\documentclass[aps,reprint,showpacs
]{revtex4-1}

\usepackage{graphicx}
\usepackage{amsfonts}
\usepackage{amssymb} 
\usepackage{thmbox} 
\usepackage{empheq} 
\usepackage{subfigure}
\usepackage{textcomp}
\usepackage{makeidx}     
\usepackage{enumerate}
\usepackage{amsmath}
\usepackage{fancyhdr}

\newcommand{\vecr}{{\vec{r}}} 

\newcommand{\lc}{{\mbox{$\mbox{\boldmath ${e}$}$}}} 

\usepackage[usenames,dvipsnames]{color}
\newcommand{\black}[1]{{\color{black}{#1}\normalcolor}}

\newcommand{\beq}{\begin{equation}}     \newcommand{\eeq}{\end{equation}}
\newcommand{\beqa}{\begin{eqnarray}}    \newcommand{\eeqa}{\end{eqnarray}}
\newcommand{\bde}{\begin{description}}  \newcommand{\ede}{\end{description}}
\newcommand{\ben}{\begin{enumerate}}    \newcommand{\een}{\end{enumerate}}

\newcommand{\la}{\langle}               \newcommand{\ra}{\rangle}

\newcommand{\ten}[1]{{\stackrel{\leftrightarrow}{#1}}}

\newcommand{\eqn}[1]{\beq{ #1 }\eeq}

\newcommand{\inv}[1]{{\frac{1}{#1}}}

\newcommand{\inRbracket}[1]{{\left({#1}\right)}}
\newcommand{\inSbracket}[1]{{\left[{#1}\right]}}

\newcommand{\veps}{{\vec{\epsilon}}} 

\newcommand{\new}[1]{{\black{#1}}}

\newcounter{formulaire}
\newcommand{\beqf}{\addtocounter{formulaire}{1}\begin{equation}}
\newcommand{\eeqf}{\tag{R \arabic{formulaire}}\end{equation}}
\newcommand{\beqaf}{\addtocounter{formulaire}{1}\begin{equation}\begin{array}{rcl}}
\newcommand{\eeqaf}{\end{array}\tag{R \arabic{formulaire}}\end{equation}}

\begin{document}


\title{Mesoscopic formulas of linear and angular momentum fluxes}


\author{Antoine Fruleux$^{1}$}
\email[Corresponding author: ]{fruleux@phare.normalesup.org}
\author{Ken Sekimoto$^{2,3}$}\email[]{ken.sekimoto@espci.fr}
\affiliation{$^1$Department of Physics and Astronomy, The University of Sheffield, Sheffield S3 7RH, United Kingdom}
\affiliation{$^2$Mati\`{e}res et Syst\`{e}mes Complexes, CNRS-UMR7057, Universit\'e   Paris-Diderot, 75205 Paris, France}  
\affiliation{$^3$Gulliver, CNRS-UMR7083, ESPCI, 75231 Paris, France}

\date{This is \jobname --- \today}

\begin{abstract}
Many approaches of coarse-graining have been developed under the names of Cosserat theory or polar-fluid theory, for those materials in which some component elements undergo non-affine deformations, such as elastic materials with inclusions or granular matters. 
For the complex elements such as living cells, however, the microscopic variables and their dynamics are often unknown, and there have been no systematic theory of coarse-graining from the \new{micro}scales, nor the formulas like Irving-Kirkwood formula that constitutes the macroscopic stress or couple-stress in terms of some \new{micro}scale quantities. 
We show that, for the quasi-steady states, the coarse-graining procedure must generally provides with  the Cosserat-type balance equations as long as the procedure keeps track of the conservation of linear and angular momenta, and that the fluxes of these conserved quantities should generally be expressed in the Irving-Kirkwood-type formulas, where the inter-particle distance or forces/torques should be replaced by those associated to the pair of neighboring coarse-graining volumes.
This framework, which refers to no particular micro-variables or dynamics, is valid for active complex matters out of equilibrium and with any multi-body interactions. 

\end{abstract}


 \maketitle


\section{Introduction\label{sec:intro}}

The Cosserat media \cite{Cosserat-Cosserat-1909}, or the micro-polar fluid \cite{Eringen-Kafadar-1976}, is a continuum description of fluid beyond the standard theories of elasticity or hydrodynamics \cite{landau-elasticity,landau-hyd} in the sense that the former description contains a characteristic scale that reflects the {mesoscopic} non-affine deformation of the constituent materials. 
The non-affine nature at small scales is reflected by explicit inclusion of angular momentum flux and also the accompanied anti-symmetric part of the momentum flux. 
When we focus on the macroscopically quasi-static processes,
the Cosserat theory represents
 the local balance of forces and torques 
by
$
  \nabla\cdot\ten{G} =\vec{0} \mbox{ and }
 \nabla\cdot\ten{{C}} =-\lc:\ten{G}, $  
where $\ten{G}$ and $\ten{C}$ are, respectively, the macroscopic momentum flux and macroscopic angular momentum flux.
$\lc$ is the Levi-Civita pseudo-tensor and the product ``$:$" is such that $(\lc:\ten{A})_\alpha=\sum_\beta\sum_\gamma\lc_{\alpha\beta\gamma}A_{\beta\gamma}$ in the cartesian component representation. 

Historically, one of the most known microscopic expressions 
for the macroscopic momentum flux, or for the stress tensor, 
is called Irving-Kirkwood (IK) formula, also called the virial stress formula \cite{Irving-Kirkwood-JCP1950,Tsai-JCP1979}. 
   Their formalism dealt with the molecules in liquid phase interacting through binary interaction potential. 
 The IK formula reads $\ten{G} = (\rho z/2)\la \vec{r}_{i,j}\otimes \vec{f}_{i,j}\ra,$ where
 the inter-particle distance $\vec{r}_{i,j}$ and the inter-particle potential force, $\vec{f}_{i,j},$ 
 constitute a tensor multiplied by the number density of the particle pairs, $\rho z/2,$ and $\la \cdot \ra$ denotes the average over the coarse-graining scale. 
\new{($\vec{X}\otimes \vec{Y}$ denotes the tensor whose $\alpha\beta$ component is $X_\alpha Y_\beta$.)} 
   There have been many extensions done of this formula, for example, to inelastic grains \cite{Babic-IntJEngSci1997}, and to polyatomic 
   constituent molecules    \cite{Dahler-Theo-AdvChemPhys1975}. 
 For the angular momentum flux $\ten{C}$ a similar formula has been derived 
 from the equations of motion, where the inter-particle force $\vec{f}_{i,j}$ is 
 replaced by the inter-particle torque $\vec{m}_{i,j}.$
For example, in \cite{Dahler-Theo-AdvChemPhys1975} $C_V$  in the second last equation in Sec.II.B
corresponds to $\ten{C}.$

The recent interests in the systems of densely packed active elements such as amoeba cells
motivate the extension of the theoretical frameworks connecting the microscopic to macroscopic scales. 
In those complex systems the cells contact among them through extended and dynamic interfaces,  and
the interactions among them naturally contain the more than two body interactions, where the `body' means the individual cell.
For such systems we  often don't know  well-defined microscopic dynamical variables 
or the microscopic model behind the non-equilibrium force-generation and responses. 
Even if we knew them,  they  could be very complicated. 
Under these circumstances, we would have difficulties to derive the macroscopic dynamics directly from very microscopic models. It would, therefore, be desirable to have 
a systematic approach that %
tells {\it (i)}  what is the canonical form of the flux balance equations at the {\it macro}\/scopic level, and {\it (ii)} what quantities at the {\it meso}\/scopic scale \new{(i.e. the cell scale)} should be given to calculate those macroscopic fluxes.
In short a {\it new} toolbox  is needed.
Once such basic toolbox is established, the remaining task is to give a suitable model for those {mesoscopic} quantities, using either those general constraints imposed by spatio-temporal symmetry and causality \cite{Cosserat-Cosserat-1909,Truesdell-Noll-1965}, or some \new{\null} simple models  as was done for the granular materials \cite{Lippman-AppMechRev1995}.

{In the present paper, we show a toolbox that will meet the above scenario. 
We show the two things: First,  whenever the conservation laws of linear and angular momenta  are correctly observed, the Cosserat-{\it type} equations are the \new{unique} relations connecting 
the macroscopic linear momentum flux $\ten{G}$ and angular momentum flux $\ten{C}$
 up to the lowest order of the small ratio between the coarse-graining scale and the characteristic scale over which the macroscopic fluxes vary. 
Secondly,  with whatsoever microscopic entities and their dynamics, 
the macroscopic fluxes $\ten{G}$ and $\ten{C}$ are expressed by the IK-{\it type} formula which act
as \new{general} mapping rules from the {mesoscopic} flow rates of momentum $\vec{F}$ and of angular momentum $\vec{M}$ to the above fluxes, $\ten{G}$ and $\ten{C},$ respectively. 
More concretely,  the momentum flux $\ten{G}$ takes the form
$\ten{G}=(\rho Z/2)\la \veps_{i,j}\otimes \vec{F}_{i,j}\ra$, where 
$\veps_{i,j}$ and $\vec{F}_{i,j}$ are, respectively, the center-to-center distance and the momentum flow rate between the pair of neighboring volumes of coarse-graining, or {\it ``cells''}, and $\rho Z/2$ is the density of such pairs in the unit volume.  
Likewise the flux $\ten{C}$ takes the form $\ten{C}=(\rho Z/2)\la \veps_{i,j}\otimes \vec{M}_{i,j}\ra$, where $\vec{M}_{i,j}$  is the angular momentum flow rate between the neighboring coarse-graining ``cells''. 
Although \new{the above} formulas look nothing but the original IK formula, it is not the case: The ``cells'' with respect to which we measure $\veps_{i,j}$, $\vec{F}_{i,j}$ and $\vec{M}_{i,j}$ are the hypothetical spatial domains adapted to our purpose of coarse-graining, while the conventional IK formula refers to the specific particles. Our formalism applies both to the discrete particle systems and the continuum ones which are either passive or active. The implication of these differences is explained in the final section in the context of the living cellular aggregates.
Throughout this paper we discuss only the quasi-static case where the inertia effects are negligible.
}

The organization of the paper is as follows. 
In the next section (\S~\ref{sec:1}) we introduce the coarse-graining ``cells'' 
 and the microscopic balances of momentum flux%
 , then we derive the balance equations of linear and angular momentum flows at the ``cell'' level 
.
In {\S}\ref{sec:3} we introduce what we call the {\it neighbor distribution function}, $\hat{\rho}_{2FM}$  (Eq. (\ref{def:rho2FC})), which characterizes the
{mesoscopic} geometry and momentum flows on the packing of coarse-graining ``cells''.
In {\S}\ref{sec:4} we first justify the replacement of the empirical distribution function, 
$\hat{\rho}_{2FM},$ by its statistical average at each spatial position%
.
Then we derive our main results, or the new toolbox mentioned above%
.
The Cosserat-type balance equations for the {macroscopic} momentum and angular momentum fluxes,  $\ten{G}$ and ${\ten{C}},$ will be derived under the IK-type definition of these
fluxes.
\new{In the concluding section \S\ref{sec:5}, after summarizing our work, we mention very briefly about an application of our framework to the aggregate of living cells, {\it Dictyostelium Discoideum} without going into details of the modeling of $\hat{\rho}_{2FM}.$ 
The Appendix provides with some details of the calculations in the main text.
}

\new{A remark is in order to avoid confusions about the terminology. We use in this paper the word {\it mesoscopic} to mention those quantities that characterize the ``cell'' interfaces such as the flow rates, $\vec{F}_{i,j},\vec{M}_{i,j},$  as well as the ``cell-cell'' distance, $\veps_{i,j}$, while in the original IK formula the corresponding quantities,
$\vec{f}_{i,j}$ and $\vecr_{i,j}$, are microscopic, whose (Newtonian) dynamics is explicitly specified.
This usage of the word {\it mesoscopic} would be a common one in the community of soft materials, which deals with the scales around $\mu\rm m$ as mesoscopic ones.
On the other hand, the community of kinetic theory might reserve this word for those quantities associated to the statistically self-averaging volume (the $\Omega$ discussed later in \S~\ref{subsec:rho-CG}), such as 
$\rho,$ $Z,$ or the two-body distribution functions. In the latter context the above-mentioned flow rates could be called sub-mesoscopic quantities.}

\section{balance laws at Microscopic level and at mesoscopic, {``cell"}, level \label{sec:1}}

\subsection{Medium and ``cell'' \label{subsec:medium}}

Our framework does not start with microscopic
 dynamical models but with the introduction of 
the {\it meso}\/scopic scale over which we \new{integrate} the momentum flux.
As for the microscopic scale we only assume the presence of the quasi-static microscopic momentum flux which will be introduced later (\S\S~\ref{subsec:2.2}). We regard the medium of our interest as a hypothetical three-dimensional packing of the closed compartments which we shall call {``cells"}. The packing can be 
disordered and slightly inhomogeneous in size.
 The typical scale of the ``cells'' should be chosen according to the modeling facility; if the material is an aggregate of amoeba, its constituent (true) cell could be chosen as ``cell''. Or if a group of molecules or grains maintains an identity of cluster, the ``cell'' could be this cluster.
The only condition on the ``cell'' is that its typical size is well below the characteristic spatial scale at which the macroscopic state varies (\S\S~\ref{sec:4}).
Up to the end of \S\S~\ref{sec:3}, however, all the descriptions below are general and 
the size of the ``cells" is arbitrary.
{Hereafter, we will write simply cell instead of ``cell'' unless there is a risk of confusion.}

We distinguish each cell by an index, $i$. 
{We assume that there is no interior free surfaces of these cells, unlike packed granular media.}
We denote by  $\vec{r}_i$ the center of volume of the $i$-th cell.
We also denote by $\Omega_i$ the space occupied by the $i$-th cell. The border of this volume, $\partial\Omega_i,$ needs not to be of polyhedral shapes, and the number of the immediate neighbor cells for each cell need not to be the same for all the cells.

\begin{figure}[h!!]
\centering
\includegraphics[scale=.5,angle=-90]{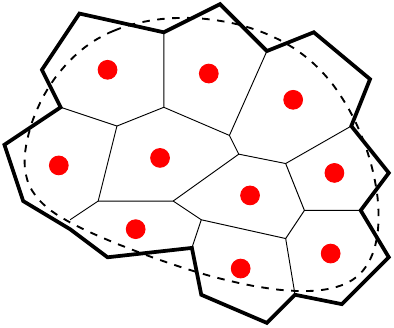}
\caption{Schematic definition of $\tilde{\Omega}$ (bounded by thick lines) for a given domain $\Omega$ (bounded by dashed curve). Cells are represented by 2D polygons and their centers inside $\Omega$ are {marked by the thick dots}. Thin lines are the ``internal'' cell-cell interfaces which do not belong to the border of $\tilde{\Omega}.$ In reality the cells are three-dimensional and the cell-cell interfaces may not be flat.}
\label{fig:Omegas}
\end{figure}

To any spatial domain $\Omega,$ either large or small,
we can associate its ``closure'' domain, 
which we denote by  $\tilde{\Omega},$ as the union of $\Omega_i$ of those cells whose center belong to $\Omega.$ In equation, it reads 
\[\tilde{\Omega}\equiv\bigcup_{\vec{r}_i\in \Omega}\Omega_i,\]
or schematically it looks like Fig.~\ref{fig:Omegas}.

\subsection{Microscopic description of the momentum balances
\label{subsec:2.2}
}

We denote by $\ten{\mathcal{G}}$  the microscopic momentum flux tensor. 
It is defined such that $\ten{\mathcal{G}}\cdot d\vec{A}(\vecr)$ gives the flow rate of momentum 
across the oriented infinitesimal surface element, $d\vec{A}(\vecr),$ at the position $\vecr$.
{We ignore completely the convective parts the linear or angular momenta which are carried by the inertia or moment of inertia, respectively.} {(We assume the existence of such microscopic description with microscopic but finite spatio-temporal resolutions.)}
Our basic starting point is the conservation laws of momentum and angular momentum,
\eqn{\label{eq:Gmicro}
\nabla\cdot\ten{\mathcal{G}}{}^t=0,\qquad 0=\lc:\ten{\mathcal{G}}{}^t,
}
 where ${}^t$ means to take the transposition. \new{\null}
The second equation is nothing but the symmetry, $\ten{\mathcal{G}}=\ten{\mathcal{G}}{}^{t},$ but we wrote above in the form similar to the final Cosserat form summarized in the final section. 
In either form, it means that the description by $\ten{\mathcal{G}}$ is enough detailed that the balance of angular momentum can be expressed by 
the linear momentum flux alone \cite{landau-elasticity}, that is, the torque on each surface element $d\vec{A}(\vec{r})$ is ignorable.
{In the setup thus defined, the quasi-static conservation of the momentum and angular momentum over the $i$-th cell reads:}
\eqn{\label{eq:gcons}
\int_{\vec{r}\in\partial\Omega_i}\ten{\mathcal{G}}\cdot d{\vec{A}}(\vec{r})=0}
\eqn{\label{eq:rgcons}
\int_{\vec{r}\in\partial\Omega_i}\vec{r}\wedge\ten{\mathcal{G}}\cdot d{\vec{A}}(\vec{r})=0,}
where integral is done over the whole boundary $\partial\Omega_i$ of the $i$-th cell occupying the volume $\Omega_i,$ and 
$d{\vec{A}}(\vec{r})$ is the outward area element at the position $\vec{r}(\in \partial \Omega_i).$
{We used the Gauss' theorem of integration to derive (\ref{eq:gcons}) and (\ref{eq:rgcons})
 from (\ref{eq:Gmicro}).}

\subsection{Coarse-graining of momentum flux --- 
``Cell''-level description of momentum balances}\label{subsec:2.3}

We rewrite the above conservation laws { (\ref{eq:gcons}) and (\ref{eq:rgcons})
  into a reduced form to the packing of the cells.}
In other words we move from the space of $\vec{r}$ to the space of the indices of the cells, $\{i\}$.
\paragraph*{Definition of inter-cellular force $F_{i,j}$ and inter-cellular torque $\vec{M}_{i,j}$ across the cell-boundary: }  
When the $i$-th  and  $j$-th cells share an interface, $\partial \Omega_i\cap\partial \Omega_j(\neq \emptyset)$, then the force and torque that the $i$-th cell applies to the $j$-th cell through this interface, which we denote by $\vec{F}_{i,j}$  and $\vec{M}_{i,j},$ respectively, are given by 
\eqn{\label{eq:defFij}
\vec{F}_{i,j}\equiv\int_{\vec{r}\in\partial\Omega_i \cap \partial\Omega_j}\ten{\mathcal{G}}\cdot d\vec{A}_{i\rightarrow j}(\vec{r})
}  
{
\eqn{\label{eq:defCij}
\vec{M}_{i,j}\equiv\int_{\vec{r}\in\partial\Omega_i \cap \partial\Omega_j}
{[\vec{r}-(\vec{r}_i+\vec{a}_{i,j})]}
\wedge\ten{\mathcal{G}}\cdot d\vec{A}_{i\rightarrow j}(\vec{r})
,
}
}
 \eqn{\label{eq:def-aij}
 \vec{a}_{i,j}=\frac{
 \int_{\partial\Omega_i \cap \partial\Omega_j}(\vec{r}-\vec{r}_i)d{A}_{i\to j}(\vec{r})}{
 \int_{\partial\Omega_i \cap \partial\Omega_j}  d{A}_{i\to j}(\vec{r})},
}
where the surface integral is done over the interface, $\partial\Omega_i \cap \partial\Omega_j,$ with $d\vec{A}_{i\rightarrow j}(\vec{r})$ being the area element at $\vec{r},$ oriented
from the $i$-th cell toward the $j$-th cell. \begin{figure}[h!!]
\centering
\includegraphics[scale=.4,angle=-90]{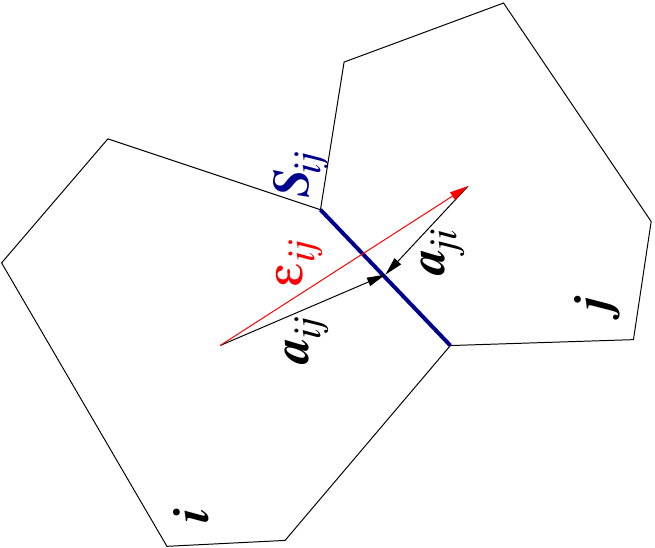}
\caption{Definition of $\vec{a}_{i,j}$ and $\veps_{i,j}$.
The common interface $\partial\Omega_i\cap \partial\Omega_j$ is denoted by $\vec{S}_{ij}$.
{The center of $S_{ij}$ is written as $\vecr_i+\vec{a}_{i,j}$ and also as $\vecr_j+\vec{a}_{j,i},$ 
where $\vecr_i$ and $\vecr_j$ are the center positions of the cell $i$ and cell $j$, respectively.}}
\label{fig:aij}
\end{figure}
The vector $\vec{a}_{i,j}$ is the relative position from $\vec{r}_i$ to the areal {\it center} of the interface,  $\partial\Omega_i \cap \partial\Omega_j,$ see Fig.~\ref{fig:aij}.
In this figure the center-to-center vector $\veps_{i,j}$ is also {introduced} as 
\eqn{\label{eq:def-eij} \veps_{i,j}=\vecr_j-\vecr_i, }  {which satisfies
the geometrical relation,  $\vec{a}_{i,j}-\vec{a}_{j,i}=\veps_{i,j},$ and
will eventually take place of $\vec{a}_{i,j}$ and $\vec{a}_{j,i}$ in the final results.
}
We can  see that the torque $\vec{M}_{i,j}$ in (\ref{eq:defCij})
is measured with respect to 
{$\vec{r}_i+\vec{a}_{i,j}$, i.e., } 
the center of the interface, ${\partial\Omega_i \cap \partial\Omega_j}.$ 
While the apparent $\vec{r}_i$ in (\ref{eq:defCij}) and that in (\ref{eq:def-aij}) cancel with each other, we retain them  since the physical meaning of the torque $\vec{M}_{i,j}$ is clearer in this form. 
{The integrals over each interface in (\ref{eq:defFij}) and (\ref{eq:defCij}) 
realize \new{the first step of} the coarse-graining of momentum by  
masking all the detailed informations in $\ten{\mathcal{G}}$ except for its zeroth ($\vec{F}$) and 
 first ($\vec{M}$) moments {while keeping track of} the conserved nature of the linear and angular momentum. }

\paragraph*{Reciprocity relations for $\vec{F}_{i,j}$ and $\vec{M}_{i,j}$: }  
\black{\mbox{} }
{The definitions (\ref{eq:defFij}) and (\ref{eq:defCij}) 
together with 
the trivial geometrical identity, $d\vec{A}_{i\rightarrow j}(\vec{r})+ d\vec{A}_{j\rightarrow i}(\vec{r})=\vec{0},$ lead to the following reciprocity relations 
about the interface between the neighboring cell-pair, $i$ and $j$:}
\beqa  \label{eqs:recipro}
\vec{F}_{i,j}+\vec{F}_{j,i}=\vec{0}\label{eq_rec},\qquad
\vec{M}_{i,j}+\vec{M}_{j,i}=\vec{0}\label{eq_Arec}.
\eeqa
{These express the conserved nature of momentum flux, or Newton's third law, across  the mesoscopic cell-cell interface.}

\paragraph*{Kirchhoff-type laws for $\vec{F}_{i,j}$ and $\vec{M}_{i,j}$ : }
\black{\mbox{} }
{The conserved nature of momentum flux 
can be also expressed for each coarse-graining cell. 
The definitions (\ref{eq:defFij}) and (\ref{eq:defCij}) with the 
simple identity, $\partial \Omega_i=\cup_j^{}(\partial\Omega_i\cap\partial \Omega_j),$  
allows to rewrite  the balance laws,  (\ref{eq:gcons}) and (\ref{eq:rgcons}), into 
the following Kirchhoff-type laws:}
\beq \label{eq:sumF}
\sum_j^{(i)} \vec{F}_{i,j}=\vec{0},
\eeq
\beq \label{eq:sumC}
\sum_j^{(i)} \left(\vec{M}_{i,j}+\vec{a}_{i,j}\wedge\vec{F}_{i,j}\right)=\vec{0},
\eeq
 where the sum $\sum_{j}^{(i)}$ runs over all the cells indexed by $j$ for which the $i$-th cell is an immediate neighbor. 
{
Notice that the flows $\vec{F}_{i,j}$ or $\vec{M}_{i,j}$ in (\ref{eqs:recipro})
or 
(\ref{eq:sumF}) and (\ref{eq:sumC}) 
do not imply exclusively the two-body interactions between the cell pairs.
  }

\paragraph*{ {
Identities for an arbitrary domain $\Omega$ : } 
}
{
Using each coarse-graining cell as a building block, the Kirchhoff-type law for a single cell can be extended to any closure domain, $\tilde{\Omega}$, that is
\footnote{
We remark that, if we defined $\vec{a}_{i,j}$ as $\veps_{i,j}/2$ (and $\vec{a}_{j,i}$ by $\veps_{j,i}/2$) instead of  (\ref{eq:def-aij}), such replacement would not affect at all the final result.
Moreover, this choice simplifies the calculation: For example, the r.h.s. of the second equation of (\ref{eq:Gtot}), which can be also written as $ \sum_{\vec{r}_i\in \Omega}\sum_j^{(i)} [\vec{M}_{i,j}+\frac{1}{2}\left(\vec{a}_{i,j}-\vec{a}_{{j,i}}\right)\wedge\vec{F}_{i,j}]{+BC}$ with the term $+BC$ representing the correction terms from the border of $\tilde{\Omega},$  should remains the same with $+BC=0$ if the above definition for $\vec{a}_{i,j}$ were taken. (Note the aforementioned identity, $\vec{a}_{i,j}-\vec{a}_{j,i}=\veps_{i,j}$) Despite this advantage,  we shall maintain the original definition of $a_{i,j}$  by  (\ref{eq:def-aij}), or equivalently by Fig.~\ref{fig:aij}, because the physical meaning of the mesoscopic torque $\vec{M}_{i,j}$ will be clearer under this definition.
}:
}
\beqa \label{eq:Gtot} 
&&\int_{\vec{r}\in\partial\tilde{\Omega}}\ten{\mathcal{G}}\cdot d{\vec{A}}(\vec{r})
=
\sum_{\vec{r}_i\in \Omega}\left\{\sum_j^{(i)} \vec{F}_{i,j}\right\},
\cr &&
\int_{\vec{r}\in\partial\tilde{\Omega}}\vec{r}\wedge\ten{\mathcal{G}}\cdot d{\vec{A}}(\vec{r})
=
\sum_{\vec{r}_i\in \Omega}\left\{\sum_j^{(i)} \left(\vec{M}_{i,j}+\vec{a}_{i,j}\wedge\vec{F}_{i,j}\right)\right\},
\cr &&
\eeqa
{where  the integrals over the cell-cell interface inside of $\tilde{\Omega}$ 
on the left hand side cancel among them. On the right hand side such cancellation is implied by the reciprocity properties given by (\ref{eqs:recipro}).}

\mbox{}

\section{Empirical ``neighbor distribution function''\label{sec:3}}

The next and crucial step is to rewrite (\ref{eq:sumF}) and (\ref{eq:sumC})
for a cell $\Omega_i$ 
in the forms which are more adapted to the {continuous field representation.
The sums in (\ref{eq:sumF}) and (\ref{eq:sumC}) contain delicate cancellations which are
closely associated to the reciprocity, (\ref{eqs:recipro}). In order to get rid of such
cancellations, }
the new idea, to our knowledge, is to use a local ``neighbor distribution function'', 
$\hat{\rho}_{2FM}$.

\paragraph*{Definition of  neighbor distribution function, $\hat{\rho}_{2FM}$: }  
This is an empirical and simultaneous distribution function of 
the center distance $\veps$, the cell-to-cell force $\vec{F}$ and cell-to-cell torque $\vec{M}$ associated to a cell at the position $\vecr$ {and its {\it immediate} neighbor. }
\begin{widetext}
\beq
\label{def:rho2FC}
\hat{\rho}_{2FM}\left(\vec{\epsilon},\vec{F},\vec{M},\vec{r}\right)
\equiv\sum_{i}\sum_{j}^{(i)}
\delta\left(\veps-\veps_{i,j}  
\right)\delta\left(\vec{F}-\vec{F}_{i,j} \right)\delta\left(\vec{M}-\vec{M}_{i,j} \right)
\delta\left(\vec{r}-\vec{r}_i \right).
\eeq
\end{widetext}
\mbox{}
{(The word ``empirical'' is used in the sense of the particular sample of statistical ensemble, and} throughout this paper, we will use the ``hat'' symbol like $\hat{A}$ to denote those empirical quantities which are {\it before} the {statistical averaging.}) 
For the simplicity of notations, we will also use the ``peripheral'' or partially integrated neighbor distribution functions, such as 
{\small$\hat{\rho}_{2F}(\vec{\epsilon},\vec{F},\vec{r})
\equiv\sum_{i}\sum_{j}^{(i)}\delta(\vec{F}-\vec{F}_{i,j} )$$\delta\left(\vec{r}_i-\vec{r} \right)$$\delta\left(\vec{r}_j-\vec{r}_i-\vec{\epsilon}\right),$}
and
\small{$\hat{\rho}_{2M}(\vec{\epsilon},\vec{M},\vec{r})
\equiv\sum_{i}\sum_{j}^{(i)}$$\delta(\vec{M}-\vec{M}_{i,j} )$$\delta\left(\vec{r}_i-\vec{r} \right)$\\
$\delta\left(\vec{r}_j-\vec{r}_i-\vec{\epsilon}\right).$}
Also we will use the purely geometrical neighbor distribution,
$\hat{\rho}_2\left(\vec{\epsilon},\vec{r}\right)
\equiv\sum_{i}\sum_{j}^{(i)}\delta\left(\vec{r}_i-\vec{r}\right)\delta\left(\vec{r}_j-\vec{r}_i-\vec{\epsilon}\right).$
The further integration of $\hat{\rho}_2(\veps,\vecr)$ over $\veps$ yields
$\int  \hat{\rho}_2\left(\vec{\epsilon},\vec{r}\right)d^3\veps=(\sum_{j}^{(i)}1)
\sum_{i}\delta\left(\vec{r}_i-\vec{r}\right),$
where {multiplicative factor,} $\sum_{j}^{(i)}1,$ is the number of neighbors of the $i$-th cell, 
{and the remainder defines} the empirical single-cell density function, $\hat{\rho}_1$ :
\beq \label{eq:rho1} 
\hat{\rho}_1\left(\vec{r}\right)
\equiv\sum_{i}\delta\left(\vec{r}_i-\vec{r}\right).
\eeq

The neighbor density distribution, $\hat{\rho}_{2FM},$ contains the detailed informations about $\veps_{i,j}=\vec{r}_j-\vec{r}_i$, $\vec{F}_{i,j}$ and $\vec{M}_{i,j}.$ 
{This function should be well distinguished from any weighting function that is introduced for the purpose of smoothing.
The right hand sides of  (\ref{eq:Gtot}) can be rewritten as moment integrals of $\hat{\rho}_{2FM}$.
Leaving the derivation to Appendix~\ref{app:A}, the result reads}
\beq \label{eq:Gbare} 
\int_{\vec{r}\in\partial\tilde{\Omega}}\ten{\mathcal{G}}\cdot d{\vec{A}}(\vec{r})
=   
\int_{\vec{r}\in \Omega}\inSbracket{
{\int\!\!\!\int\!}\vec{F}\, \hat{\rho}_{2F}(\vec{\epsilon},\vec{F},\vec{r})\,d^3\vec{F}\,d^3\vec{\epsilon}}\,d^3\vec{r},
\eeq
{\small
\beqa \label{eq:rGcook}
&&\int_{\vec{r}\in\partial\tilde{\Omega}}\vec{r}\wedge\ten{\mathcal{G}}\cdot d{\vec{A}}(\vec{r})
\cr &&=
\int_{\vec{r}\in \Omega}\!\! \inSbracket{
{\int\!\!\!\int\!\!\!\int\!} 
(\vec{M}\,+\frac{1}{2}\vec{\epsilon}\wedge\vec{F})
\hat{\rho}_{2FM}(\vec{\epsilon},\vec{F},\vec{M},
\vec{r})\,d^3\vec{\epsilon}\,d^3\vec{M} \,d^3\vec{F} } d^3 \vec{r}.
\cr &&
\eeqa
}
\paragraph*{Redundancy of $\hat{\rho}_{2FM}(\vec{\epsilon},\vec{F},\vec{M},\vec{r})$: }
\black{\mbox{} }
{
By looking at the  pair of neighboring cells at $\vec{r}$ and at $\vec{r}+\veps$ in two ways,
the reciprocity relations (\ref{eqs:recipro}) are expressed as 
a redundancy property of the neighbor distribution function : 
}
\eqn{  \label{eq_redun}
\hat{\rho}_{2FM}\left(\vec{\epsilon},\vec{F},\vec{M},\vec{r}\right)=
\hat{\rho}_{2FM}\left(-\vec{\epsilon},-\vec{F},-\vec{M},\vec{r}+\vec{\epsilon}\right)
.}
{The derivation is given in Appendix~\ref{app:B}, but 
the contents may be intuitively understandable.}
The associated peripheral distributions {inherit the similar relations} :
\beq \label{eq_rec2}
\hat{\rho}_{2F}\left(\vec{\epsilon},\vec{F},\vec{r}\right)=\hat{\rho}_{2F}\left(-\vec{\epsilon},-\vec{F},\vec{r}+\vec{\epsilon}\right),
\eeq
\beq \label{eq_Arec2}
\hat{\rho}_{2M}\left(\vec{\epsilon},\vec{M},\vec{r}\right)=\hat{\rho}_{2M}\left(-\vec{\epsilon},-\vec{M},\vec{r}+\vec{\epsilon}\right).
\eeq
{The redundancy relation implies the following identities, whose derivations are given in
Appendix~\ref{app:C}:}
\beqa \label{cor:Frho2F}
&&{\int\!\!\!\int\!} \vec{F}\hat{\rho}_{2F}\left(\vec{\epsilon},\vec{F},\vec{r}\right)\,d^3\vec{F}\,d^3\vec{\epsilon}
\cr &&=
\inv{2}{\int\!\!\!\int\!} \vec{F}\,\inSbracket{
\hat{\rho}_{2F}\left(\vec{\epsilon},\vec{F},\vec{r}\right)-
\hat{\rho}_{2F}\left(\vec{\epsilon},\vec{F},\vec{r}-\vec{\epsilon}\right)}
\,d^3\vec{F}\,d^3\vec{\epsilon}
\cr &&
\eeqa  
\beqa \label{eq:eeFhat}
&&{\int\!\!\!\int\!}\vec{\epsilon} \otimes \vec{\epsilon} \otimes \vec{F}\hat{\rho}_{2F}\left(\vec{\epsilon},\vec{F},\vec{r}\right)\,d^3\vec{F}\,d^3\vec{\epsilon}
\cr &&
+{\int\!\!\!\int\!} \vec{\epsilon} \otimes \vec{\epsilon} \otimes\vec{F}\hat{\rho}_{2F}\left(\vec{\epsilon},\vec{F},\vec{r}-\vec{\epsilon}\right)\,d^3\vec{F}\,d^3\vec{\epsilon}
=0
\eeqa
{
These identities will be used below (\S~\ref{sec:4}) to get rid of the delicate cancellations in the Kirchhoff-Type laws, (\ref{eq:sumF}) and (\ref{eq:sumC}).}

{
\section{Macroscopic fluxes of momentum and angular momentum
\label{sec:4}} 
}

{
\subsection{Statistically averaged neighbor distributions\label{subsec:rho-CG}}
}

{As long as the size of the coarse-graining cells is 
properly chosen (see \S\S~\ref{subsec:medium}),
the {\it statistical} properties of the neighbor distribution function $\hat{\rho}_{2FM}$ 
should be mostly homogeneous over the spatial domain $\Omega$ whose diameter 
is well above the former size but, at the same time, well below the characteristic spatial scale at which the macroscopic state varies.
When a  function including  $\hat{\rho}_{2FM}$ as a factor is integrated
over the whole $\veps$-, $\vec{F}$- and $\vec{M}$- domains but limited over the  \new{$\vecr$-}domain $\Omega$,
we can replace  $\hat{\rho}_{2FM}$ by its statistical average as a good approximation thanks to the law of large number.
We will denote by ${\rho}_{2FM}$ without    ``$\,\,\hat{\mbox{}}\,\,$''  the statistical average of 
$\hat{\rho}_{2FM} .$
We will use this averaged distribution function with the understanding that all the macroscopic properties
can be calculated through the integral over the above mentioned semi-macroscopic domain $\Omega$
(or its closure $\tilde{\Omega}$), see for example (\ref{eq:Gbare}) and (\ref{eq:rGcook}).
The statistically averaged version of the peripheral distribution functions, $\rho_{2F}(\veps,\vec{F},\vec{r}),$ ${\rho}_{2M}(\veps,\vec{M},\vec{r}),$ $\rho_{2}(\veps,\vec{r})$ and  $\rho_{1}(\vec{r}),$ are also introduced. Under the same premise,} the empirical number of cell neighbors, $\hat{Z}$, is replaced by its statistical average, $Z(\vec{r}),$ which is defined through 
\eqn{ \label{eq:meanZ}   
Z (\vec{r})\rho_1(\vec{r})
\equiv \int{\rho_2\left(\vec{\epsilon},\vec{r}\right)}d^3\vec{\epsilon}.
}
{The statistically averaged neighbor distribution function, ${\rho}_{2FM},$ and its peripheral distributions  should be slowly varying functions in space, $\vecr$. They, therefore, allow the gradient expansion in $\veps,$ that is, the spatial derivative operation with respect to $\vecr$,  which we denote by $\nabla$, satisfies the property,} $\|\veps\cdot\nabla\|\ll 1,$ where $\|\veps\|$ is the typical center-to-center distance of neighboring cell pairs. {We will use this expansion below.}

\mbox{}
{
\subsection{Expression for the macroscopic fluxes\label{subsec:macro}}
}

{
In the integral conservation relations, (\ref{eq:Gbare}) and (\ref{eq:rGcook}), 
we apply the redundancy relations  (\ref{eq_rec2}) and (\ref{eq_Arec2}), %
in the way shown in (\ref{cor:Frho2F})  or its homologous form with $\vec{F}$ and $\vec{M}$ being exchanged. Then we replace the empirical distributions by the statistical averaged ones to 
apply the gradient expansion mentioned above. The higher order terms in the expansion is much smaller than the first order, which we can verify using the identity (\ref{eq:eeFhat}).
Leaving the details of derivation in Appendices~\ref{app:D} and \ref{app:E},
the results read as follows.

For the macroscopic} momentum flux $\ten{G},$ {it} satisfies the local momentum conservation,
\beq \label{eq_consG}  \nabla\cdot\ten{G} =\vec{0}. \eeq
{under the definition:}
\eqn{\label{def:tenG}
\ten{G}(\vecr)\equiv
\frac{Z(\vecr)\rho_1(\vecr) }{2}\la  \veps
\otimes\vec{F} \ra_2 (\vecr)  ,
}
where  
$\la\,\mathcal{\psi}\ra_2(\vecr)$ stands for the average of $\mathcal{\psi}$ {
with the statistically averaged} neighbor distribution function {${\rho}_{2FM}$}, 
\beqa\label{eq:av2def}
\la \mathcal{\psi} \ra_2(\vecr)
&\equiv &
\frac{
{\int\!\!\!\int\!\!\!\int\!} \mathcal{\psi}\, \,
{\rho}_{2FM}\left(\vec{\epsilon},\vec{F},\vec{M},\vec{r}\right)\,d^3\vec{M}\,d^3\vec{F}\,d^3\vec{\epsilon}
}{
{\int\!\!\!\int\!\!\!\int\!}  {\rho}_{2FM}\left(\vec{\epsilon},\vec{F},\vec{r}\right)\,d^3\vec{M}\,d^3\vec{F}\,d^3\vec{\epsilon}
}
\cr &=& \frac{{\int\!\!\!\int\!\!\!\int\!} \mathcal{\psi}\, \,
{\rho}_{2FM}\left(\vec{\epsilon},\vec{F},\vec{M},\vec{r}\right)\,d^3\vec{M}\,d^3\vec{F}\,d^3\vec{\epsilon}}{Z (\vec{r})\rho_1(\vec{r})}.
\eeqa

{For the macroscopic} angular momentum flux $\ten{{C}},$ {it} satisfies
\beq  \label{eq_consC} \nabla\cdot\ten{{C}} =-\lc:\ten{G}, \eeq
{under the definition:}
\eqn{\label{def:mathcalC}
\ten{{C}}(\vecr)\equiv 
{
\frac{Z(\vecr)\rho_1(\vecr) }{2}\la  \veps
\otimes\vec{M} \ra_2(\vecr) }.
}
\new{While the Eqs. (\ref{def:tenG}) and (\ref{def:mathcalC}) used
the factorization of $Z(\vecr)$ and $\rho_1(\vecr)$ to conform with the original IK formula, 
what are to calculate is the moments of $\rho_{2FM}(\veps,\vec{F},\vec{M},\vecr)$, such as
${\int\!\!\!\int\!} \vec{\epsilon}\otimes\vec{F}\, \rho_{2F}\left(\vec{\epsilon},\vec{F},\vec{r}\right)\,d^3\vec{F}\,d^3\vec{\epsilon}$ in Eq.(\ref{eq:533gives}).
Under a quasi-uniformly packed compartments, $\{\Omega_i\}$,
the factors $Z(\vecr)$ and $\rho_1(\vecr)$  are mostly uniform and constant. The relevant macro-variables are rather the flows of mass, momentum and angular momentum and some order parameters such as the mean polarity of the cellular units.}
\black{\mbox{ } }\\

\section{Conclusion and discussion \label{sec:5}}
{We recognized that, 
in the media undergoing the quasi-steady processes, 
(i) the Cosserat-type equations provide with the general principle of macroscopic characterization of the linear and angular momenta, 
(\ref{eq_consG}) and (\ref{eq_consC}), 
and that (ii) the Irving-Kirkwood-type or virial-type formulas provide with the general principle of transformation of these momenta from mesoscopic to macroscopic level,
(\ref{def:tenG}) and (\ref{def:mathcalC}),
where $\vec{F}$ and $\vec{M}$ are, respectively, the flow rate of linear and angular momenta
that pass through the hypothetical interface between a pair of coarse-graining cells with
$\veps$ being their center-to-center distance.
The average $\la\cdot \ra_2(\vec{r})$ defined by  (\ref{eq:av2def}) essentially samples
over many pairs of coarse-graining cells around the given position $\vec{r}$ with $Z(\vecr)\rho_1(\vecr)/2$ being the spatial density of such pairs.
\new{For those complex and non-equilibrium systems whose micro-variables and dynamics are unknown or too complicated, it is a matter of choice whether we take a completely phenomenological approach to correlate {\rm directly} the macro flux variables to the macro state variables (e.g. \cite{ArctiveGel-Prost-PRL2004,ActiveGel-Prost-EPJE2005}),  or, otherwise, we take an {\rm indirect} approach that takes advantage of the statistical mechanics, which is our choice. 
}
}

{ 
Our results are distinguished 
 from the original Irving-Kirkwood (IK) formulas:
We relate the macroscopic fluxes to those flows of momentum and angular momentum which are attributed to each {\it mesoscopic} cells, or volume elements, over which we carry out \new{our first step of} coarse-graining. Unlike the conventional scheme the variables such as distance vectors, forces or torques among the {\it microscopic} model elements (i.e., the molecules or grains) do not appear there. 
{\it Our IK-type formulas consist of, instead of  the inter-particle force or torque,  the linear and angular momentum flows across the interface separating the neighboring coarse-graining cells and, instead of the inter-particle distance, there appear the distance between the centers of these cells.}  
Although our theory is also a ``bottom-up'' construction, we coarse-grain the mesoscopic momentum flux instead of any microscopic {\it dynamical model}. 
 The coarse graining without micro-variables or dynamics is possible because we deal with just as many components for the two conserved vector fields to derive the two vectorial conservation laws.
 
 Once the correct IK-type formulas for the fluxes are given, 
they naturally satisfy the Cosserat-type equations whatever is the microscopic ingredients. 
This logic is consistent with the ``microscopic'' Cosserat equations in
\cite{Eringen-Kafadar-1976} (the Eqs. (1.4.12) and (1.4.13) in Part I), where the authors
obtained as an equivalent form of the balance equations of momentum and angular momentum 
{\it without coarse-graining} but with a continuum field theory of {\it polar} fluid containing the microscopic angular momentum density (``spin'').

Our work thus reveals the universality of Cosserat type equations and Irving-Kirkwood type formulas through an abstract procedure of the coarse-graining.
Although being much in narrower sense than the Newton's laws or the laws of thermodynamics as general principles,
we assert that the Cosserat type equations and Irving-Kirkwood type formulas are still the principles that apply to a wide class of phenomena than were ever discussed.
In these principle, 
the size of the coarse-graining cell could be different from that of the constituent individual elements of the system when, for example, those elements form local clusters.
}

{
An important outcome of our finding 
is its applicability to the systems of densely packed active elements such as amoeba cells. In fact it was such system that first motivated us to develop the present framework. 
\new{As mentioned in \S \ref{sec:intro}, the interaction and dynamics of such system is usually very  complicated.}
Even under such circumstances our theory can still tell that 
(i) the Cosserat-type equations in terms of the fluxes, $\ten{G}$ and $\ten{C}$, are the canonical representation of the flux balance at the {\it macro}\/scopic level 
and that (ii) it is the flows $\vec{F}$ and $\vec{M}$ associated to the coarse-graining cell interfaces 
that should be found in order to calculate $\ten{G}$ and $\ten{C}.$
Our framework thus provides a {\it new} toolbox to those domains of research 
which have not been exploited by 
 the kinetic theories of gases and granular matters or the theory of continuum with inclusions.
}
\begin{figure}[h]
\centering
\subfigure[\null ] 
{   \label{fig:hard1sur2}  \includegraphics[width=9cm]{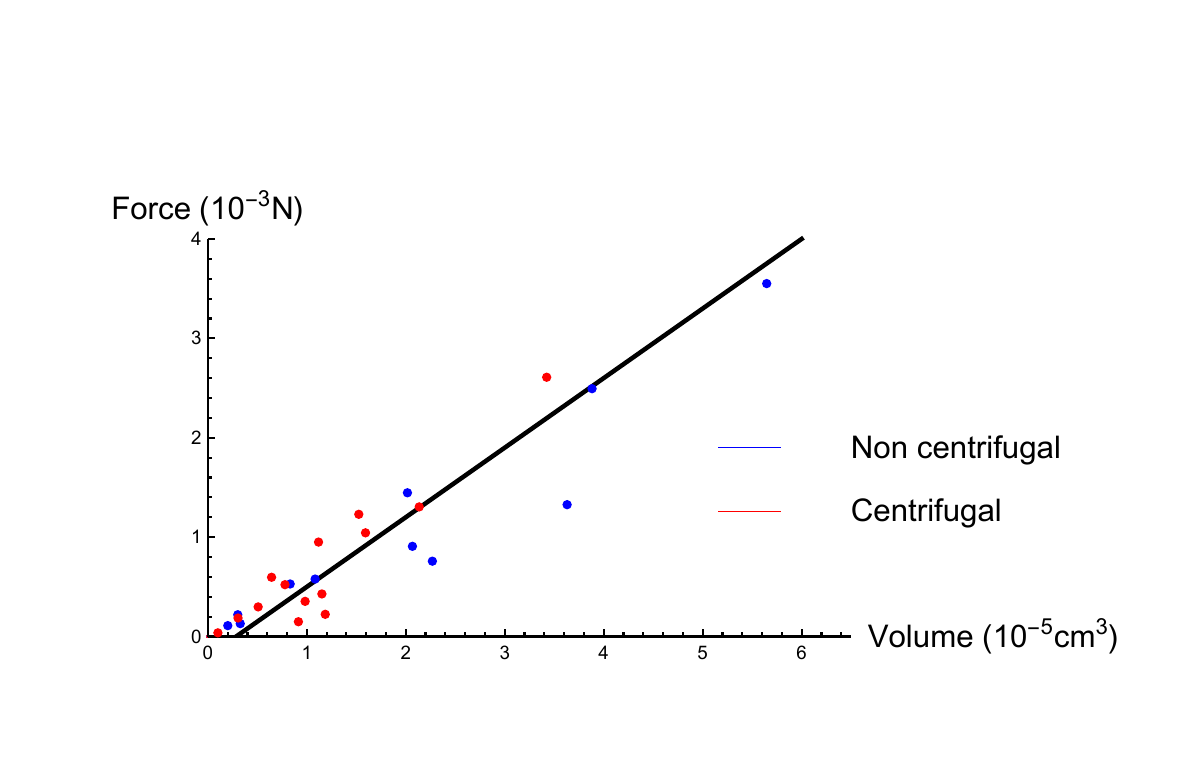}}
\hspace{0.5cm}
\subfigure[\null ] 
{   \label{fig:hard2sur3}  \includegraphics[width=6cm]{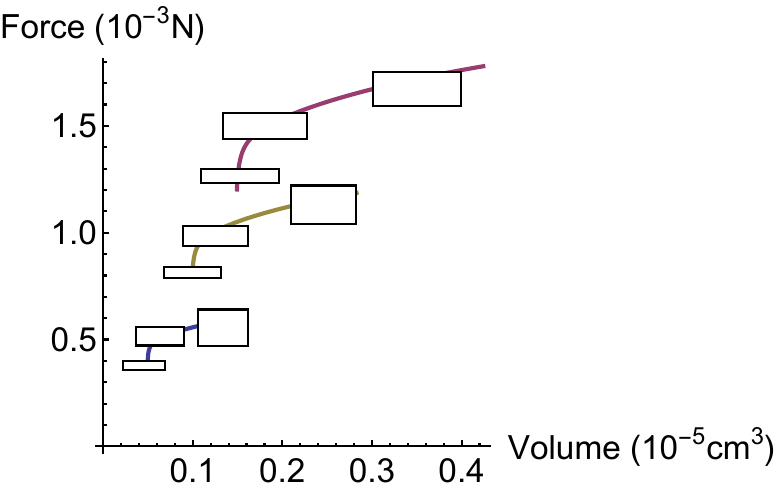}}
\caption{
\protect\new{(a) The experimental stall forces generated by an aggregate of {\it Dictyostelium Discoideum} when its directed collective movement in a tube is impeded either by a pressure gradient (blue dots, ``Non centrifugal'') or by a centrifugal force (red, ``Centrifugal''). The data are extracted from Inoue {\it et al.}  (\protect\cite{VolumicForce-Inouye-JCS1980} Fig.~5 and  \protect\cite{VolumicForceCentrfuge-Inouye-Protoplasma1984} Fig.~5).
(b) Model prediction of the stall force using the present framework for the macroscopic momentum and angular momentum fluxes. The results are reproduced from \protect\cite{fruleux:tel-01095839}  Fig.~9.14. The model is two-dimensional and the rectangular symbols represent the shape and dimensions of the aggregates. Those aggregates with the same length along the tube axis are grouped by the continuous curves. 
 }}
\label{fig:Hard} 
\end{figure}

%
 In order to apply to specific setup, we need to supplement with proper boundary conditions adapted to the characteristic of the system. Unlike the conventional hydrodynamics of viscous fluids, the no-slip boundary condition is not {\it a priori} assured and we should carefully choose the appropriate ones case by case. 

\new{%
Although the case studies are 
not our purpose in the present paper, we briefly introduce below
a comparison between the theory using the present framework \cite{fruleux:tel-01095839} 
and the experiments done about the force generation by the aggregate of {\it Dictyostelium Discoideum} \cite{Gilbert}. 
The experiments have measured the stall force of the aggregate when its collective movement is impeded by an adjustable external force. The stall force was found to depend approximatively linearly on the volume of the aggregates \protect\cite{VolumicForce-Inouye-JCS1980, VolumicForceCentrfuge-Inouye-Protoplasma1984}, see Fig.~\ref{fig:Hard}(a). 
This result was intriguing because, if the locomotive cells were simply packed with an aligned polarity,
only those cells at the surface of the aggregate could contribute to the locomotive force, implying the proportionality of the stall force to the contact area with the substrate surface.
In the theory \cite{fruleux:tel-01095839} 
the effect of the local deviations of the polarity of cells in the presence of the substrate are taken into account. The resultant linear and angular momentum flows among the cells are modeled for $\vec{F}_{i,j}$
and $\vec{M}_{i,j}$ through the (pair) neighbor distribution function, $\rho_{2FM},$
with a due consideration of the boundary conditions.
{In order to evaluate $\rho_{2FM}$
the theory \cite{fruleux:tel-01095839} not only uses the known behavioral facts about the `crawling' of individual cells in a dense environment but also introduces the three-neighbor distribution function. Since the microscopic dynamics for the ``cells'' is not available, the three-neighbor distribution function is not related through BBGKY-like hierarchy to $\rho_{2FM}.$
Instead, an extensive use have been made of the redundancy properties of the former function. 
The neighbor distribution functions in our new approach, $\rho_{2FM}$, plays the central role bridging between the {\it mesoscopic} flow rates, $\vec{F}_{i,j}$ and $\vec{M}_{i,j}$, and the macroscopic rheological variables.}
The analysis of the model predicted the stall force that increases with the volume of the aggregates, see Fig.~\ref{fig:Hard}(b). In this model those cells with distorted polarity constitute the boundary layers whose thickness depends on the stall force due to the non-linearity of the system. As a result the stall force should also depends on the aspect ratio of the aggregate \cite{fruleux:tel-01095839}. The theory is free of adjustable parameters, and the quantitative discrepancy about a factor $\sim 5$ could be due to the basic parameters taken from different literatures.}

While \new{%
the main advantage of the present framework is its applicability to the materials consisting of complex mesoscopic elements,} our formalism may also shed new light on the existing theories on the gas kinetics and polar fluid dynamics (see, for example, \cite{Closure-Weinhart2012} and the references cited therein):
Many efficient methods of coarse-graining which have been developed so far, such as 
through the Chapman-Enskog theory, the moment integrals 
\cite{Eringen-Kafadar-1976,Dahler-Theo-AdvChemPhys1975, Irving-Kirkwood-JCP1950},
or integral over space-time with weighting functions  
 \cite{Babic-IntJEngSci1997} or test function \cite{Eringen-Kafadar-1976}, 
 have sometimes put more priority on some physical or practical aspects other 
than the conservation of linear- and angular-momenta. 
For example, if we use the volume or spatio-temporal integral of the microscopic forces over a coarse-graining (space-time) volume with some weighting function, the conserved nature of the microscopic momentum may not be automatically %
inherited by the resulting {coarse-grained} system. 
That is, the macroscopic fluxes might obey the equations which are not exactly of 
the Cosserat-type. In particular, the source of divergence of 
angular momentum flux ($-\lc:\ten{G}$ for the Cosserat-type) would be different from the antisymmetric part of the momentum flux ($\ten{G}$) which appears in the conservation of the linear momentum ($\nabla\cdot\ten{G}=0$).
For example, in \cite{Dahler-Theo-AdvChemPhys1975}, 
 the structure of the Cosserat equations %
is lost because the moment integrals used there do not necessarily assure the flux conservation.  
Logically, such theory can also be correct about the momentum conservation if the modifications from the IK-type fluxes 
are exactly compensated by the modifications from the Cosserat-type equations.
Our purpose, however, is to provide with a framework that observes explicitly the momentum balance
at the mesoscopic scale so that we can focus on the modeling of the mesoscopic flows.
To our knowledge there is only a few paper that used the coarse-graining 
along this line for {\it non}-polar fluids \cite{Todd-Evans-MOP-pre1995}. There the authors used the surface integral instead of the volume integral of the momentum flux
although they focused rather on the numerical improvement than the conservation issue.

\begin{widetext}
\appendix

\section{Derivation of (\ref{eq:Gbare}) and (\ref{eq:rGcook}) \label{app:A}}
First we show 
\beq \label{eq:sumsumFij}
\sum_{\vec{r}_i\in \Omega}\left\{\sum_j^{(i)} \vec{F}_{i,j}\right\}
=
\int_{\vec{r}\in \Omega}{\int\!\!\!\int\!}\vec{F}\, \hat{\rho}_{2F}(\vec{\epsilon},\vec{F},\vec{r})\,d^3\vec{F}\,d^3\vec{\epsilon}\,d^3\vec{r}
\eeq
and
\beq \label{eq:sumsumCij}
 \sum_{\vec{r}_i\in \Omega}\left\{\sum_j^{(i)}  \vec{M}_{i,j}\right\}
=
\int_{\vec{r}\in \Omega}{\int\!\!\!\int\!}\vec{M}\, \hat{\rho}_{2M}(\vec{\epsilon},\vec{M},\vec{r})\,d^3\vec{M}\,d^3\vec{\epsilon}\,d^3\vec{r}
\eeq
for arbitrary $\Omega$.
Eq.  (\ref{eq:sumsumFij}) holds because
\beqa
\nonumber
\sum_{\vec{r}_i\in \Omega}\left\{\sum_j^{(i)} \vec{F}_{i,j}\right\} 
 &=&\int_{\vec{r}\in \Omega}\int \sum_{i}\sum_{j}\vec{F}_{i,j}\delta\left(\vec{r}_j-\vec{r}\right)\delta\left(\vec{r}_j-\vec{r}_i-\vec{\epsilon}\right) \,d^3\vec{\epsilon}\,d^3\vec{r}\cr
&=&\int_{\vec{r}\in \Omega}{\int\!\!\!\int\!} \sum_{i}\sum_{j}\vec{F}\delta\left(\vec{F}-\vec{F}_{i,j}\right)\delta\left(\vec{r}_j-\vec{r}\right)\delta\left(\vec{r}_j-\vec{r}_i-\vec{\epsilon}\right) \,d^3\vec{F}\,d^3\vec{\epsilon}\,d^3\vec{r}
\cr
&=&\int_{\vec{r}\in \Omega}{\int\!\!\!\int\!}\vec{F}\, \hat{\rho}_{2F}(\vec{\epsilon},\vec{F},\vec{r})\,d^3\vec{F}\,d^3\vec{\epsilon}\,d^3\vec{r}
\eeqa
The equality for the torque (\ref{eq:sumsumCij}) can be shown similarly.
Next, combining the above results with (\ref{eq:Gtot}) we have the identities,
\beq \label{eq:rGbare}
\int_{\vec{r}\in\partial\tilde{\Omega}}\vec{r}\wedge\ten{\mathcal{G}}\cdot d{\vec{A}}(\vec{r})
=
-\int_{\vec{r}\in \Omega}{\int\!\!\!\int\!}\vec{M}\, \hat{\rho}_{2M}(\vec{\epsilon},\vec{M},\vec{r})\,d^3\vec{M}\,d^3\vec{\epsilon}\,d^3\vec{r}
-\sum_{\vec{r}_i\in \Omega}\left\{\sum_j^{(i)} \vec{a}_{i,j}\wedge\vec{F}_{i,j}\right\}
\eeq
On the r.h.s.,
$\sum_{\vec{r}_i\in \Omega}\sum_j^{(i)} \left(\vec{a}_{i,j}\wedge\vec{F}_{i,j}\right)$
can also be represented as a moment integral: we first notice that this sum is equal to 
$
\frac{1}{2}\sum_{\vec{r}_i\in \Omega}\sum_j^{(i)}\left(\vec{a}_{i,j}-\vec{a}_{{j,i}}\right)\wedge\vec{F}_{i,j}$. Then, by the geometrical identity $\vec{a}_{i,j}-\vec{a}_{i,j}=\veps_{i,j}$,
we have $\sum_{\vec{r}_i\in \Omega}\sum_j^{(i)} \left(\vec{a}_{i,j}\wedge\vec{F}_{i,j}\right)
=\frac{1}{2}\sum_{\vec{r}_i\in \Omega}\sum_j \veps_{i,j}\wedge\vec{F}_{i,j},$ which is written as
$\frac{1}{2}\int_{\vec{r}\in \Omega}\inRbracket{
{\int\!\!\!\int\!} \vec{\epsilon}\wedge\vec{F}\, \,\rho_{2F}\left(\vec{\epsilon},\vec{M},\vec{r}\right)\,d^3\vec{\epsilon}\,d^3\vec{M} }d^3 \vec{r}.$
Therefore, (\ref{eq:rGbare}) reads (\ref{eq:rGcook}).
{\it (End of derivation)}

\section{Derivation of (\ref{eq_redun}) \label{app:B}}
In the definition of $\hat{\rho}_{2FM}$ (see (\ref{def:rho2FC}))
\[
\hat{\rho}_{2FM}\left(\vec{\epsilon},\vec{F},\vec{M},\vec{r}\right)
\equiv\sum_{i}\sum_{j}^{(i)}\delta\left(\vec{r}_j-\vec{r}_i-\vec{\epsilon}\right)\delta\left(\vec{F}-\vec{F}_{i,j} \right)\delta\left(\vec{M}-\vec{M}_{i,j} \right)\delta\left(\vec{r}-\vec{r}_i \right)
\]
we change the argument as follows;
\beqa
\hat{\rho}_{2FM}\left(-\vec{\epsilon},-\vec{F},-\vec{M},\vec{r}+\vec{\epsilon}\right)
&=&
\sum_{i}\sum_{j}^{(i)}\delta\left(\vec{r}_j-\vec{r}_i-\vec{\epsilon}\right)\delta\left(\vec{F}+\vec{F}_{i,j} \right)\delta\left(\vec{M}+\vec{M}_{i,j} \right)\delta\left(\vec{r}+\vec{\epsilon} -\vec{r}_i\right)
\cr &=&
\sum_{i}\sum_{j}^{(i)}\delta\left(\vec{r}_j-\vec{r}_i-\vec{\epsilon}\right)\delta\left(\vec{F}+\vec{F}_{i,j} \right)\delta\left(\vec{M}+\vec{M}_{i,j} \right)\delta\left(\vec{r} -\vec{r}_j\right).
\eeqa 
Using the fact that the counting of all the immediate neighbors through the sum,
$\sum_{i}\sum_{j}^{(i)},$ is equivalent to that through
$\sum_{j}\sum_{i}^{(j)}$, the last line becomes
\[
\hat{\rho}_{2FM}\left(-\vec{\epsilon},-\vec{F},-\vec{M},\vec{r}+\vec{\epsilon}\right)
=
\sum_{j}\sum_{i}^{(j)}\delta\left(\vec{r}_j-\vec{r}_i-\vec{\epsilon}\right)\delta\left(\vec{F}+\vec{F}_{i,j} \right)\delta\left(\vec{M}+\vec{M}_{i,j} \right)\delta\left(\vec{r} -\vec{r}_j\right)
\]
Finally, the reciprocity relations, 
$\vec{F}_{i,j}=-\vec{F}_{j,i}$ and $\vec{M}_{i,j}=-\vec{M}_{j,i}$
gives 
\beqa
\hat{\rho}_{2FM}\left(-\vec{\epsilon},-\vec{F},-\vec{M},\vec{r}+\vec{\epsilon}\right)
&=&
\sum_{j}\sum_{i}^{(j)}\delta\left(\vec{r}_j-\vec{r}_i-\vec{\epsilon}\right)\delta\left(\vec{F}-\vec{F}_{j,i} \right)\delta\left(\vec{M}-\vec{M}_{j,i} \right)\delta\left(\vec{r} -\vec{r}_j\right)
\cr &=& 
\hat{\rho}_{2FM}\left(\vec{\epsilon},\vec{F},\vec{M},\vec{r}\right).
\eeqa
We thus arrived at the basic redundancy relationship (\ref{eq_redun}) claimed above.   
The associated relations for the peripheral distributions can be obtained by integrating over either   $\vec{M}$ or $\vec{F}$.
  {\it (End of derivation.)}

\section{Derivation of (\ref{cor:Frho2F}) and (\ref{eq:eeFhat}) \label{app:C}} 
For  (\ref{cor:Frho2F}),
 \beqa
 {\int\!\!\!\int\!} \vec{F}\hat{\rho}_{2F}\left(\vec{\epsilon},\vec{F},\vec{r}\right)\,d^3\vec{F}\,d^3\vec{\epsilon}
 &&= 
\inv{2} \inSbracket{
{\int\!\!\!\int\!} \vec{F}\hat{\rho}_{2F}\left(\vec{\epsilon},\vec{F},\vec{r}\right)\,d^3\vec{F}\,d^3\vec{\epsilon}+{\int\!\!\!\int\!} \vec{F}\hat{\rho}_{2F}\left(-\vec{\epsilon},-\vec{F},\vec{r}+\vec{\epsilon}\right)\,d^3\vec{F}\,d^3\vec{\epsilon}
}
\cr &&= 
\inv{2} \inSbracket{
{\int\!\!\!\int\!} \vec{F}\hat{\rho}_{2F}\left(\vec{\epsilon},\vec{F},\vec{r}\right)\,d^3\vec{F}\,d^3\vec{\epsilon}+{\int\!\!\!\int\!} (-\vec{F})\hat{\rho}_{2F}\left(\vec{\epsilon},\vec{F},\vec{r}+(-\vec{\epsilon})\right)\,d^3\vec{F}\,d^3\vec{\epsilon}
}
\cr &&= 
\inv{2}{\int\!\!\!\int\!} \vec{F}\,\inSbracket{
\hat{\rho}_{2F}\left(\vec{\epsilon},\vec{F},\vec{r}\right)-
\hat{\rho}_{2F}\left(\vec{\epsilon},\vec{F},\vec{r}-\vec{\epsilon}\right)}
\,d^3\vec{F}\,d^3\vec{\epsilon}.
\eeqa
For (\ref{eq:eeFhat})
\beqa 
{\int\!\!\!\int\!}\vec{\epsilon} \otimes \vec{\epsilon} \otimes \vec{F}\hat{\rho}_{2F}\left(\vec{\epsilon},\vec{F},\vec{r}\right)\,d^3\vec{F}\,d^3\vec{\epsilon}
 &&={\int\!\!\!\int\!} \vec{\epsilon} \otimes \vec{\epsilon} \otimes\vec{F}\hat{\rho}_{2F}\left(-\vec{\epsilon},-\vec{F},\vec{r}+\vec{\epsilon}\right)\,d^3\vec{F}\,d^3\vec{\epsilon}
\cr &&=
-{\int\!\!\!\int\!} \vec{\epsilon} \otimes \vec{\epsilon} \otimes\vec{F}\hat{\rho}_{2F}\left(\vec{\epsilon},\vec{F},\vec{r}-\vec{\epsilon}\right)\,d^3\vec{F}\,d^3\vec{\epsilon}.
\eeqa
By adding the l.h.s. and the 2nd line on the r.h.s., we have (\ref{eq:eeFhat}).
{\it (End of derivation)}

\section{Derivation of (\ref{eq_consG}) and (\ref{def:tenG})\label{app:D}} 
We look at (\ref{eq:Gbare}) and rewrite its right hand side.
Because $\mathcal{\psi}$ will not contain $\vec{M}$ in this issue, we will use $\rho_{2F}$ to simplify the notation. According to (\ref{cor:Frho2F})
 or, more precisely, to its statistically averaged version, 
\beq
\label{eq:Fhatrho2bis}
{\int\!\!\!\int\!} \vec{F}{\rho}_{2F}\left(\vec{\epsilon},\vec{F},\vec{r}\right)\,d^3\vec{F}\,d^3\vec{\epsilon}
=
\inv{2}{\int\!\!\!\int\!} \vec{F}\,\inSbracket{
{\rho}_{2F}\left(\vec{\epsilon},\vec{F},\vec{r}\right)-
{\rho}_{2F}\left(\vec{\epsilon},\vec{F},\vec{r}-\vec{\epsilon}\right)}
\,d^3\vec{F}\,d^3\vec{\epsilon}.
\eeq
The integral on the r.h.s. can be expanded using its slowly varying nature about $\vecr$:
\beqa \label{eq:533gives}
\mbox{(r.h.s. of (\ref{eq:Fhatrho2bis}))}
&=& \frac{1}{2}\nabla\cdot
\left\{{\int\!\!\!\int\!} \vec{\epsilon}\otimes\vec{F}\, \rho_{2F}\left(\vec{\epsilon},\vec{F},\vec{r}\right)\,d^3\vec{F}\,d^3\vec{\epsilon}\right\}
\cr &&+
\frac{1}{2}\nabla\nabla:\left\{{\int\!\!\!\int\!} \veps\otimes\vec{\epsilon}\otimes\vec{F}\, \rho_{2F}\left(\vec{\epsilon},\vec{F},\vec{r}\right)\,d^3\vec{F}\,d^3\vec{\epsilon}\right\}
+\mathcal{O} \left( \| \veps\cdot\nabla\,\|^3\right).
\eeqa
Now the second term on the r.h.s., which is at most $\mathcal{O} \left( \| \veps\cdot\nabla\,\|^2\right)$, is in fact only $\mathcal{O} \left( \| \veps\cdot\nabla\,\|^3\right)$ because 
the $\veps\cdot\nabla$ expansion of (\ref{eq:eeFhat}) shows that
\beqa 
2{\int\!\!\!\int\!}\vec{\epsilon} \otimes \vec{\epsilon} \otimes \vec{F}\hat{\rho}_{2F}\left(\vec{\epsilon},\vec{F},\vec{r}\right)\,d^3\vec{F}\,d^3\vec{\epsilon}
\,-\,\nabla\cdot{\int\!\!\!\int\!} \veps\otimes\vec{\epsilon} \otimes \vec{\epsilon} \otimes\vec{F}\hat{\rho}_{2F}\left(\vec{\epsilon},\vec{F},\vec{r}\right)\,d^3\vec{F}\,d^3\vec{\epsilon}
\,\simeq \,0,
\eeqa
therefore,
\beqa
\nabla\nabla:\inSbracket{
{\int\!\!\!\int\!} \veps\otimes\vec{\epsilon}\otimes\vec{F}\, \rho_{2F}\left(\vec{\epsilon},\vec{F},\vec{r}\right)\,d^3\vec{F}\,d^3\vec{\epsilon}
}
&\simeq& \nabla\nabla:\inSbracket{
\inv{2}
\nabla\cdot\left\{{\int\!\!\!\int\!} \veps\otimes\veps\otimes\vec{\epsilon}\otimes\vec{F}\, \rho_{2F}\left(\vec{\epsilon},\vec{F},\vec{r}\right)\,d^3\vec{F}\,d^3\vec{\epsilon}\right\}}
\cr &=&  \mathcal{O} \left( \| \veps\cdot\nabla\,\|^3\right).
\eeqa
Therefore, the first term  on the r.h.s. of (\ref{eq:533gives}) 
is a good approximation with an error of only $\mathcal{O} \left( \| \veps\cdot\nabla\,\|^3\right).$
Now rewriting  (\ref{eq:533gives})  
using the notation of $\la \veps\otimes\vec{F}\ra_2$ and the definition  (\ref{def:tenG}), 
the equation (\ref{eq:Fhatrho2bis})  becomes
\[
{\int\!\!\!\int\!} \vec{F}{\rho}_{2F}\left(\vec{\epsilon},\vec{F},\vec{r}\right)\,d^3\vec{F}\,d^3\vec{\epsilon}
= \nabla\cdot \ten{G}+\mathcal{O} \left( \| \veps\cdot\nabla\,\|^3\right).
\]
The (\ref{eq:Gbare}) takes the form:
$\int_{\vec{r}\in\partial\tilde{\Omega}}\ten{\mathcal{G}}\cdot d{\vec{A}}(\vec{r})
=\int_{\vec{r}\in \Omega} \nabla\cdot\ten{G} \,d^3\vec{r}.
 $
for arbitrary $\Omega$ up to the errors of $\mathcal{O} \left( \| \veps\cdot\nabla\,\|^3\right)$.
 This justifies to identify $\ten{G}$ as the {macroscopic} momentum flux.
In terms of the suffix, it is rather the transpose, $\ten{G}^t,$ that corresponds to $\ten{\mathcal{G}}$ due to our choice of representation of the latter.
Finally (\ref{eq:rgcons}) leads to (\ref{eq_consG}).
{\it (End of derivation)}

\section{Derivation of (\ref{eq_consC}) and (\ref{def:mathcalC}) \label{app:E} }
By a completely parallel argument as the derivation of the previous theorem,
a part of the integral on the r.h.s. of (\ref{eq:rGcook}) can be rewritten as:
\beq
\int_{\vec{r}\in \Omega}{\int\!\!\!\int\!}\vec{M}\, \hat{\rho}_{2M}(\vec{\epsilon},\vec{M},\vec{r})\,d^3\vec{M}\,d^3\vec{\epsilon}\,d^3\vec{r}
=   \nabla\cdot \ten{{C}} +  \mathcal{O} \left( \| \veps\cdot\nabla\,\|^3\right),
\eeq
where we have used  the definition (\ref{eq:av2def})  and (\ref{def:mathcalC}).
Besides, the remaining part of the integral in (\ref{eq:rGcook}), i.e.,\\
$-\int_{\vec{r}\in \Omega}\inSbracket{{\int\!\!\!\int\!\!\!\int\!} \frac{1}{2}\vec{\epsilon}\wedge\vec{F}
\, \, \hat{\rho}_{2FM}\left(\vec{\epsilon},\vec{F},\vec{M}, \vec{r}\right)\,d^3\vec{\epsilon}\,d^3\vec{M} \,d^3\vec{F} }d^3 \vec{r},$
can be expressed in terms of $\ten{G}$ defined in (\ref{def:tenG}):
\beqa
\int_{\vec{r}\in \Omega}\inRbracket{
{\int\!\!\!\int\!} 
\frac{1}{2}\vec{\epsilon}\wedge\vec{F}\, \,
\rho_{2F}\left(\vec{\epsilon},\vec{M},\vec{r}\right)\,d^3\vec{\epsilon}\,d^3\vec{M} }
d^3 \vec{r}
 &=&
\int_{\vec{r}\in \Omega} \inSbracket{\frac{1}{2}\la\vec{\epsilon}\wedge\vec{F} \ra_2{Z}\rho_1\, }
d^3\vec{r}
\cr &=& 
\int_{\vec{r}\in \Omega} \lc:\inSbracket{\frac{1}{2}\la\vec{\epsilon}\otimes\vec{F} \ra_2{Z}\rho_1\, }
\cr &=& 
{\int_{\vec{r}\in \Omega} \lc:\ten{G} \, d^3\vec{r}}.
\eeqa
Now, if we ignore the errors of $\mathcal{O} \left( \| \veps\cdot\nabla\,\|^3\right)$,
 (\ref{eq:rGcook}) then reads
\[
\int_{\vec{r}\in\partial\tilde{\Omega}}\vecr\wedge\ten{\mathcal{G}}\cdot d{\vec{A}}(\vec{r})
=
\int_{\vec{r}\in \Omega} (\nabla\cdot\ten{{C}}+\lc:\ten{G}) \,d^3\vec{r}
\]
for arbitrary $\Omega.$ 
This equation together with (\ref{eq:rgcons}) leads to (\ref{eq_consC}).
{\it (End of derivation)} 

\end{widetext}

%
\begin{acknowledgments}
AF was financially supported by Ecole Doctorale de Physique en \^{I}le de France.
KS thanks the laboratory Gulliver at E.S.P.C.I. for its hospitality during the achievement of the present work.
\end{acknowledgments}


\bibliography{AFbiblio-C.bib}

\end{document}